\documentclass{PoS}
\usepackage{amsmath}

\title{Exclusive production of heavy quarkonia as a probe of the low $x$ and low scale gluon PDF}

\ShortTitle{Exclusive production of heavy quarkonia as a probe of the low $x$ and low scale gluon PDF}

\author{\speaker{C.~A.~Flett}$^a$, S.~P.~Jones$^b$,  A.~D.~Martin$^c$, M.~G.~Ryskin$^{c,d}$ and T.~Teubner$^a$\\
\llap{$^a$} Department of Mathematical Sciences, University of Liverpool, Liverpool L69 3BX, U.K.\\
\llap{$^b$} Theoretical Physics Department, CERN, Geneva, Switzerland \\
\llap{$^c$} Institute for Particle Physics Phenomenology, Durham University, Durham DH1 3LE, U.K.\\
\llap{$^d$} Petersburg Nuclear Physics Institute, NRC Kurchatov Institute, Gatchina, St. Petersburg, 188300, Russia \\
E-mail: \email{c.a.flett@liverpool.ac.uk}, \email{s.jones@cern.ch}, \email{a.d.martin@durham.ac.uk}, \email{ryskin@thd.pnpi.spb.ru}, \email{thomas.teubner@liverpool.ac.uk}}

\abstract{We discuss the exclusive $J/\psi$ photoproduction process, as measured recently at LHCb, as a means of constraining and ultimately determining the low $x$ and low $Q$ gluon PDF. The scale dependence of the theoretical prediction for this process is shown to be systematically improved via a taming of the known $\overline{\text{MS}}$ result, this amounts to resumming logarithmically enhanced small-$x$ terms and implementing a small-$Q$ power correction.
The cross section level predictions allow the behaviour of the gluon PDF in the low $(x, Q)$ domain to be determined.}
\FullConference{Light Cone 2019 - QCD on the light cone: from hadrons to heavy ions - LC2019\\
		16-20 September 2019\\
		Ecole Polytechnique, Palaiseau, France}

\begin{document}

\section{Introduction}

The recent measurement by LHCb at $ \sqrt{s} = \text{13 TeV}$~\cite{Ronan} of central exclusive production of $J/\psi$ mesons in $pp \rightarrow p\,+\,\,J/\psi\,+\,p$ events allows constraints on the low $x$ and low $Q$ scale behaviour of the global gluon parton distribution function (PDF).  Together with data from HERA~\cite{H1ZEUS02} and the LHCb at a lower collision energy of $\sqrt{s}=\text{7 TeV}$ \cite{Ronan}, the kinematic coverage is extended down to about $x = (M_{J/\psi}/\sqrt{s}) e^{-Y} \sim 3 \times 10^{-6}$ with $Q \sim \text{1.5 GeV},$ and where $2<Y(J/\psi)<4.5$ is the rapidity interval of the LHCb forward detector.

We study exclusive $J/\psi$ photoproduction to NLO within collinear factorisation and show that a systematic taming of the $\overline{\text{MS}}$ approach leads to a refined theoretical result that, until recently, was plagued by large scale dependence and large logarithms at high energy (small $x$). As we will see, this is overcome through implementation of a $Q_0$ cut and resummation of a class of double logarithms that, by virtue of the low scale process, are necessary to ensure a reliable and stable theoretical prediction.

In Section 2, we briefly recall our model construction and emphasise the challenges one faces in incorporating the exclusive $J/\psi$ framework into global PDF analyses. In Section 3, we discuss the new and improved theoretical result, before mapping this to the cross section level which allows preliminary predictions about the behaviour of the low $x$ and low scale gluon PDF to be made.

\section{Model framework and challenges}

We describe the ultraperipheral event $pp \rightarrow p\,\,J/\psi\,\,p$ within collinear factorisation,
with the set up for the quasi-elastic (semi) hard scattering subprocess $\gamma p \rightarrow J/\psi\, p$ following \cite{Ivanov}. Explicitly, 
\begin{equation}
    \text{Im $A$} \sim \sum_{i=q,g} \text{Im} (F_i \otimes C_i) \otimes \phi_{Q \bar Q}^V,
\end{equation}
where the $C_i$ are the perturbatively calculable subprocess kernels, 
and the $F_i$ denote {\it generalised} parton distribution functions (GPDs). These are a generalisation of the conventional collinear PDFs used in the global analyses, required by the
off forward kinematics and the presence of skewing parameter, $\xi$, see e.g. \cite{Diehl}. The projection of the open charm quarks onto the $J/\psi$ wave function, $\phi_{Q \bar Q}^V$,  is made within LO NRQCD. Hoodbhoy \cite{Hood} showed that relativistic corrections to the $c \bar c \,\, J/\psi$ transition vertex do not amount to a sizeable correction to the cross section and would mostly affect the normalisation only which, as we will see, is adequate given the description of the data in the HERA region (where the global gluon PDF is known better). 
Relativistic corrections do not affect the behaviour as a function of energy, which is completely
driven by the $x$ dependence of the gluon PDF.

Our construction of the GPD grids is via the Shuvaev integral transform. It is a well established means of obtaining GPDs, with the non perturbative collinear PDFs as input, assuming we are at a sufficiently small $x$ and that the transform is used in the correct region of parameter space to avoid Regge pole singularities; $\xi < |X| \ll 1$, see Fig. 1 for our set up.   At NLO, the accuracy is $\mathcal O(x)$ so,  even for the slightly larger $x$ probed at HERA, the error is marginal. The imaginary part of the coefficient function describes the hard matrix element and we include the real part of the amplitude through the dispersion relation, 
$$\frac{\text{Re $A$}}{\text{Im $A$}} = \frac{\pi}{2} \frac{\partial \ln \text{Im} A/W^2}{ \partial \ln W^2},$$ where $W$ is the $\gamma p$ subprocess centre of mass energy. 
\begin{figure}[h]
    \centering
    \includegraphics[scale=0.4]{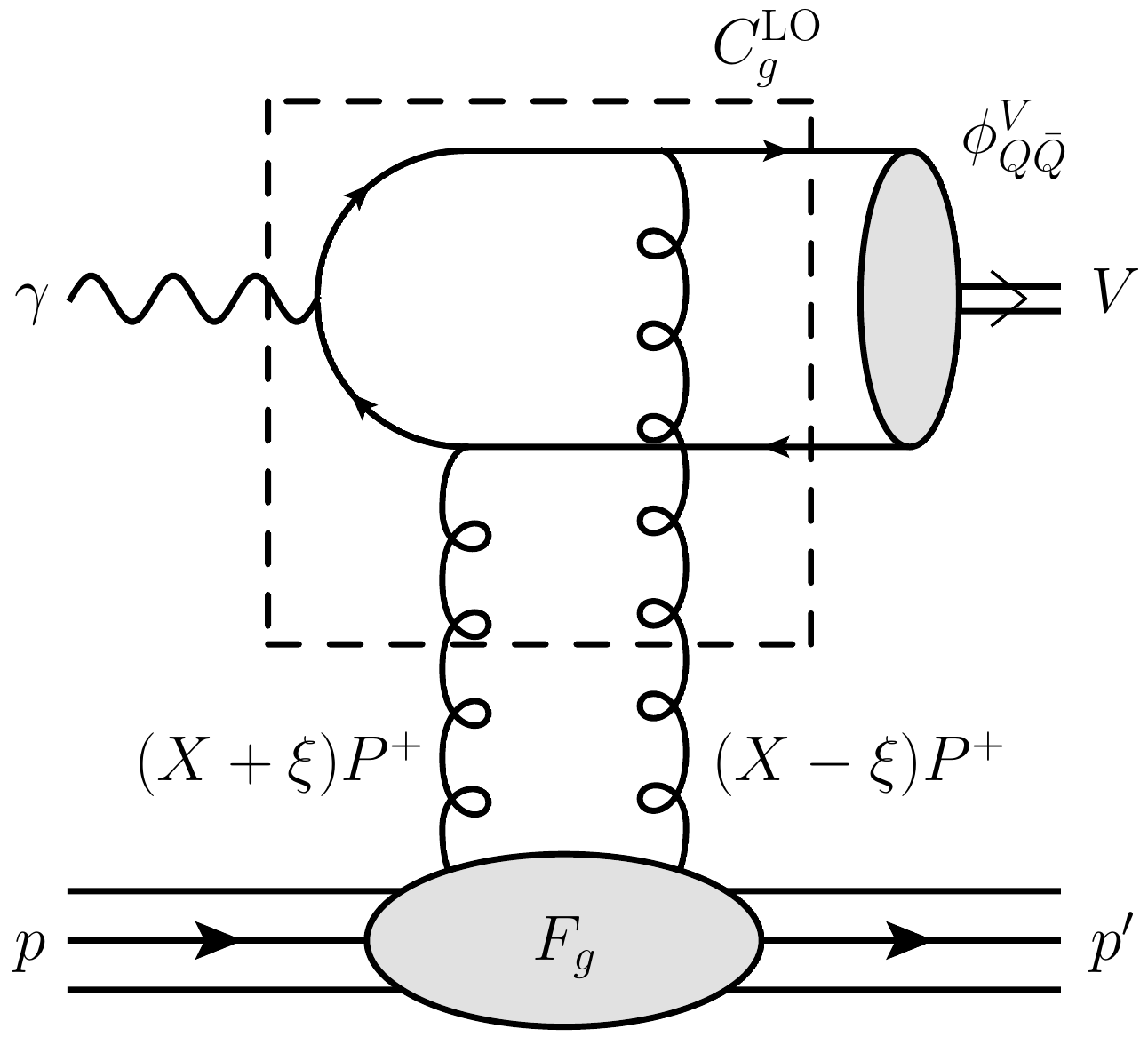}
    \qquad
    \includegraphics[scale=0.4]{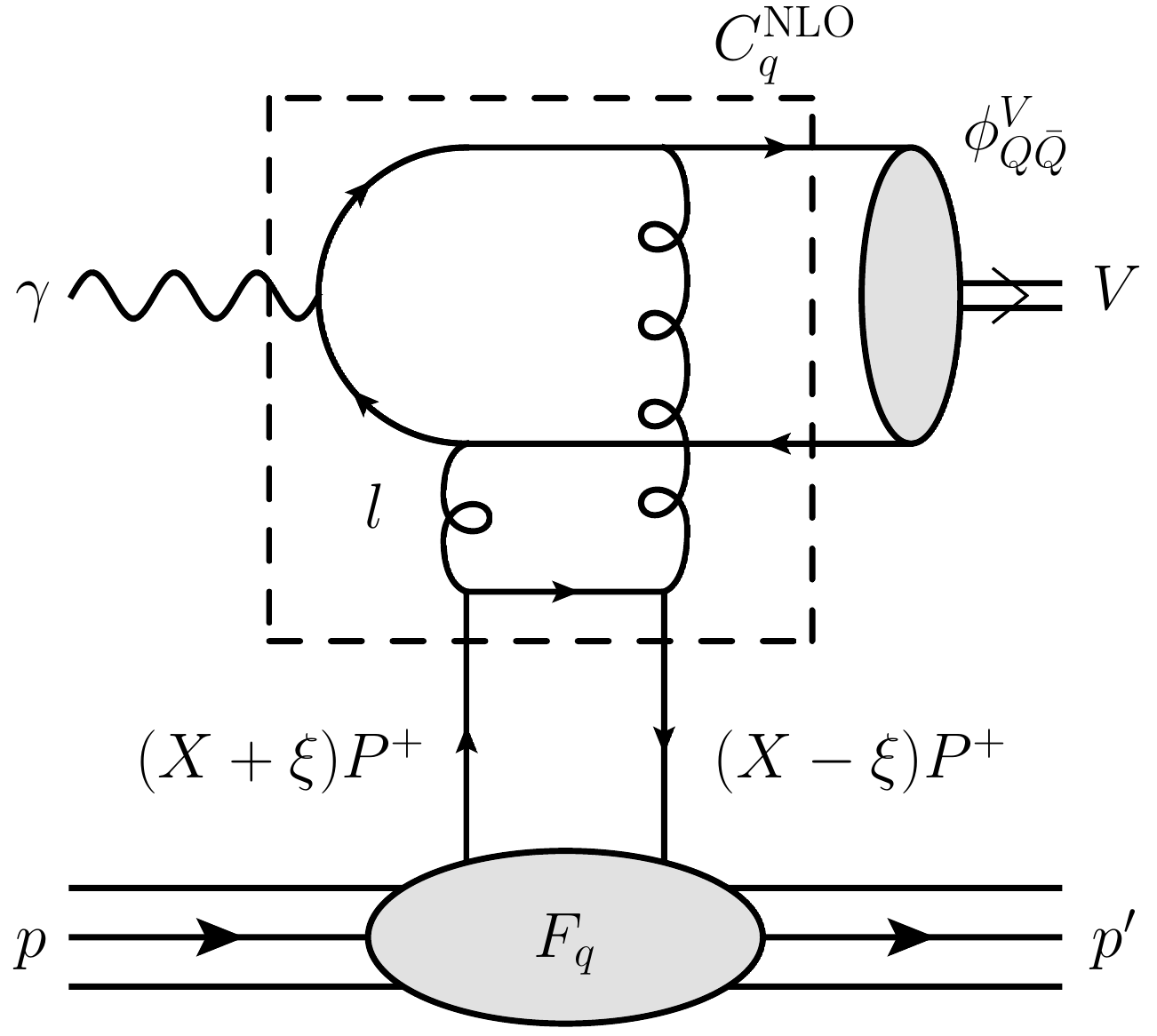}
    \caption{The LO gluon initiated (left panel) and NLO quark initiated (right panel) $J/\psi$ production diagrams. Here the $F_i$ are the GPDs, the dashed boxed encloses the perturbatively calculable five leg coefficient function and $\phi^V_{Q\bar{Q}}$ describes the formation of the $J/\psi$. } 
    \label{fig:my_label}
\end{figure}

The LHCb collaboration extract the cross section for $\gamma p \rightarrow J/\psi\,p$ using 
\begin{equation}
\frac{\text{d} \sigma^{\text{th}}(pp)}{\text{d} Y} = S^2(W_+) N_+ \sigma^{\text{th}}_+ (\gamma p) + S^2(W_-) N_- \sigma^{\text{th}}_- (\gamma p),
\end{equation} 
with $S^2(W_{\pm})$ and $N_{\pm}$ rapidity gap survival factors and photon fluxes, respectively, for $\gamma p$ centre of mass energies $W^2_{\pm} = M_{J/\psi} \sqrt{s} e^{\pm |Y|}.$ The survival factors, estimated from general rescattering principles, account for factorisation breaking corrections and describe the probability that the rapidity gap will not become populated with additional emissions that would otherwise destroy the exclusivity of the event. The $W_-$ component samples $x \sim 10^{-2}$ and so $\sigma_{-}$ may be provided by the HERA data while the $W_+$ component samples $x \sim 10^{-5}$ or less and so $\sigma_+$ can be extracted from eqn. (2.2) using $\text{d} \sigma^{\text{th}}(pp)/\text{d} Y$ provided by LHCb.

\section{Stability of amplitudes at NLO and cross section predictions}
The NLO contribution for exclusive $J/\psi$ photoproduction in the $\overline{\text{MS}}$ collinear factorisation scheme has been known for some time \cite{Ivanov}. However, the result in this scheme has been brought under control, only recently, by means of eliminating a double counting effect and resumming a class of large logarithms, see \cite{doublelog} for full details.  One may 
analyse the high energy asymptotics of the NLO amplitude and observe that the result contains terms $\sim \alpha_s(\mu_R)^2 \ln (\mu_F^2/m_c^2) \ln(1/\xi)$, which upset the perturbative convergence through a resulting large scale dependence at small $\xi$. The renormalisation and factorisation scales are denoted by $\mu_R$ and $\mu_F$ respectively. However, by setting $\mu_F = m_c$, these terms are completely absorbed providing a resummation for this class of large logarithmically enhanced terms.  

The upshot is a shifting of terms from the NLO coefficient function to the LO GPD with a remaining scale dependence, $\mu_F$, in the NLO coefficient function and a residual scale dependence, $\mu_f$, in the NLO GPD. The resulting $\gamma p \rightarrow J/\psi \,\,p$ amplitude is therefore recast in the form
\begin{equation}
    A(\mu_f) = C_g^{\text{LO}} \otimes F_g(\mu_F) + \sum_{i=q,g} C^{\text{NLO}}_{\text{rem}, i} (\mu_F) \otimes F_i(\mu_f).
\end{equation}
With this choice, we have absorbed large contributions arising  from a specific scale and momentum fraction hierarchy into
the parametrisation of the input GPD.

We must, in addition, also investigate the effect of a subtraction that is important at low scales. Recall the DGLAP evolution starts at some low, yet still perturbative, scale $Q_0$. At LO, all contributions below $Q_0$ are included in the input PDFs at $Q_0$. At NLO, however, we need to subtract from the evolution the contribution of $t$-channel loop momentum $|l|^2 < Q_0^2$,
otherwise we double count. In this way, we restrict the virtuality of the four momentum circulating in the gluon ladder diagrams to be above $Q_0$. Note that this has never been a ubiquitous feature of an $\overline{\text{MS}}$ calculation but is important as the subtraction amounts to a power correction of $\mathcal O(Q_0^2/\mu_F^2)$, which is sizeable here because the process sits at a low scale, of the order of $m_c$\footnote{Actually, it can be shown that the $Q_0$ subtraction effectively absorbs the quark contribution so that the exclusive $J/\psi$ process is predominantly a probe of the gluon PDF.}.
Fig. 2 shows the vast suppression of scale variation achieved compared to the $\overline{\text{MS}}$ approach.
\begin{figure}[h]
    \centering
    \includegraphics[scale=0.38]{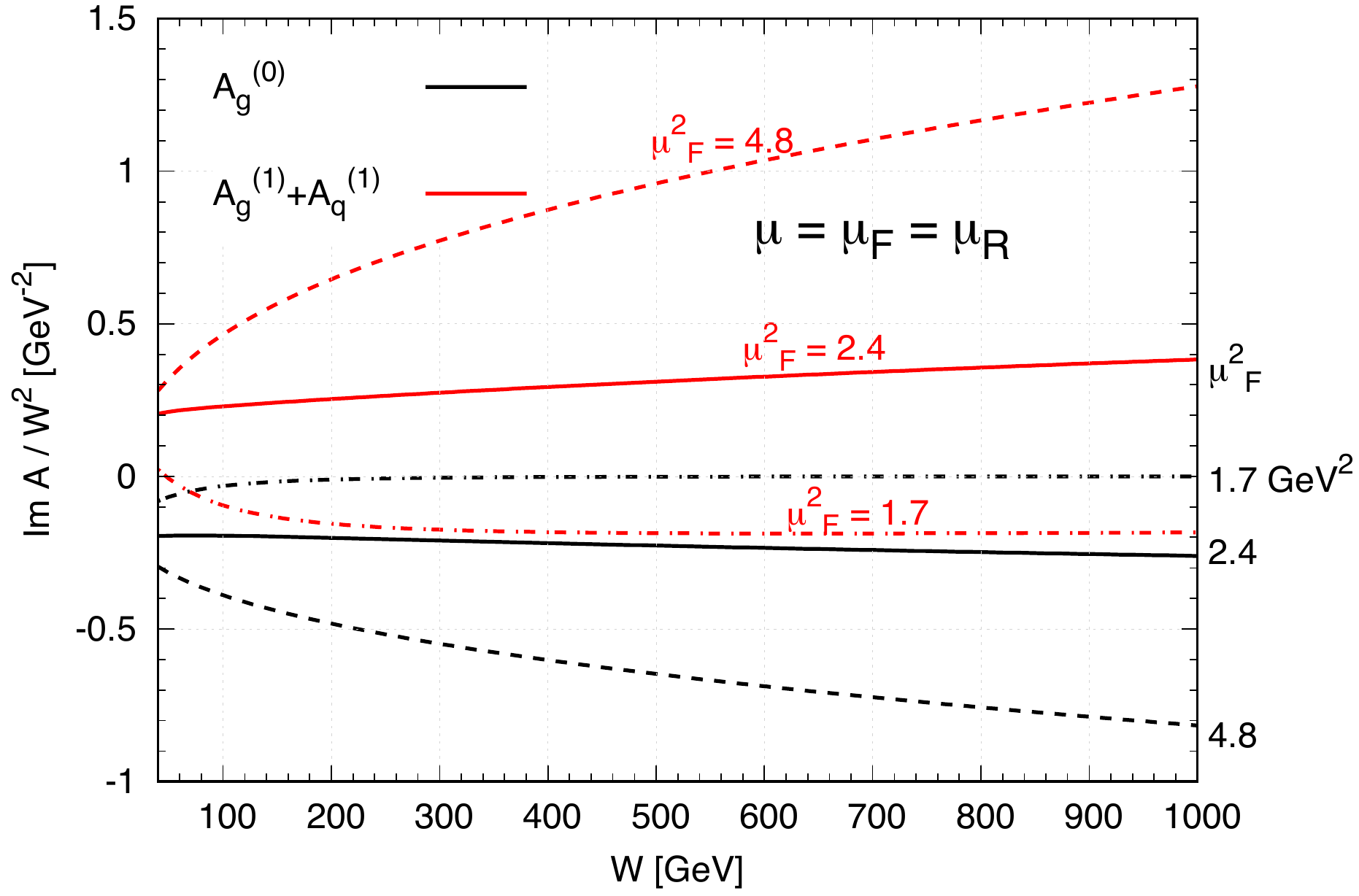}
    \qquad
    \includegraphics[scale=0.38]{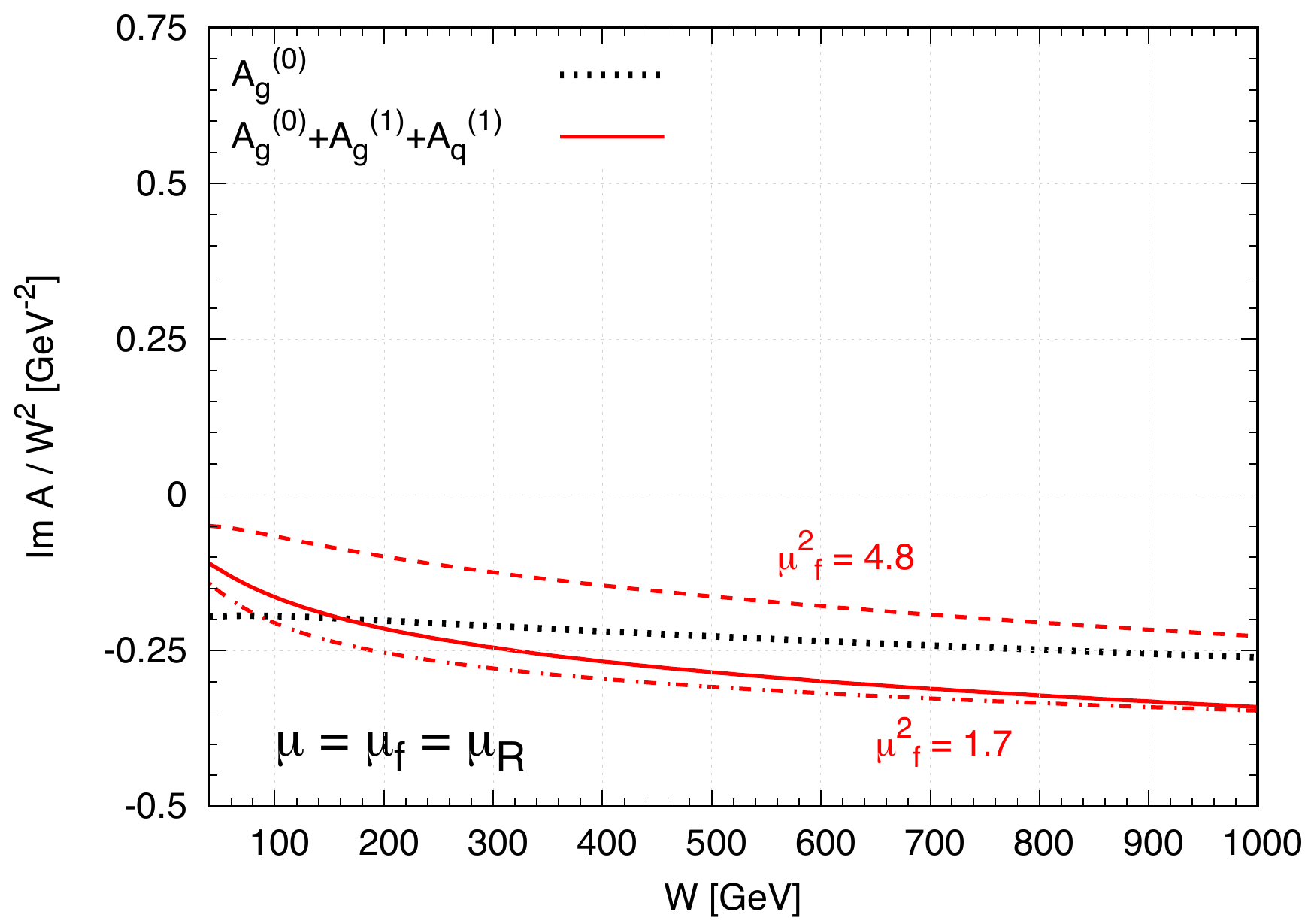}
    \caption{$\overline{\text{MS}}$ scale variations of $\text{Im }A/W^2$ vs. $W$ at LO and NLO generated using CTEQ6.6 partons at $\mu_F^2 = \mu_R^2 = Q_0^2,\, m_c^2,\, 2 m_c^2$ (left panel). The corresponding plot with now the scale fixing and $Q_0$ cut implemented, with variations now made w.r.t $\mu_f =\mu_R$ and $\mu_F^2 = m_c^2 = 2.4\, \text{GeV}^2$ fixed (right panel). $Q_0 = 1.3\, \text{GeV}$ is the starting PDF scale for CTEQ6.6. } 
    \label{fig:my_label}
\end{figure}


Might a fully fledged BFKL resummation improve the situation further? It was suggested in~\cite{28paper} to resum instead terms like $\sim \alpha_s \ln(1/\xi)$ present in the coefficient function to obtain better stability with respect to factorisation scale variations. However, we do not do so here as the $Q_0$ subtraction would need to be implemented in a manner consistent with the use of the LO BFKL kernel and, furthermore, the LO BFKL gives rise to a too hard gluon PDF, which is inconsistent with the LHCb data.

Now, with the NLO result sufficiently stable, the natural next step is to compare the predictions with the data, that is, at the level of the cross section.
In Fig. 3 (left panel), we show predictions using three sets of global PDF fits \cite{NNPDF3.0, CT14, MMHT14}, evaluated at $Q_0 = \mu_F = \mu_f = \mu_R = m_c$. The results agree favourably in the HERA regime, where the global partons are better constrained, whilst in the LHCb regime we observe huge differences between the various global PDFs\footnote{Exclusive $J/\psi$ photoproduction was also studied within the $k_t$ factorisation framework, see~\cite{kt}. Here, a DLL modified input gluon ansatz mimicking the resummation of double logarithms through one step of DGLAP evolution was used and fitted to the experimental data. In this approach, however, only a subset of the full NLO corrections are incorporated. As the tamed collinear factorisation result includes all the effects within $k_t$ factorisation (to NLO accuracy) but with the full NLO contribution realised, it is in this sense an improved theoretical prediction. Moreover, the gluon extracted in this alternative approach is not readily comparable to the $\overline{\text{MS}}$ partons of the global analyses, without a non-trivial scheme conversion. Hence, we do not compare at a quantitative level here.}. This, together with the right panel in which we propagate the individual global fit uncertainties to the cross section, shown as a shaded band for MMHT14 and boundary curves for NNPDF3.0, demonstrates the utility of the exclusive $J/\psi$ data and supports the claim that their inclusion into
the global PDF analyses would provide a serious constraint for the low $x$ and low $Q$ behaviour of the gluon PDF.
\begin{figure}[h]
    \centering
    \includegraphics[scale=0.395]{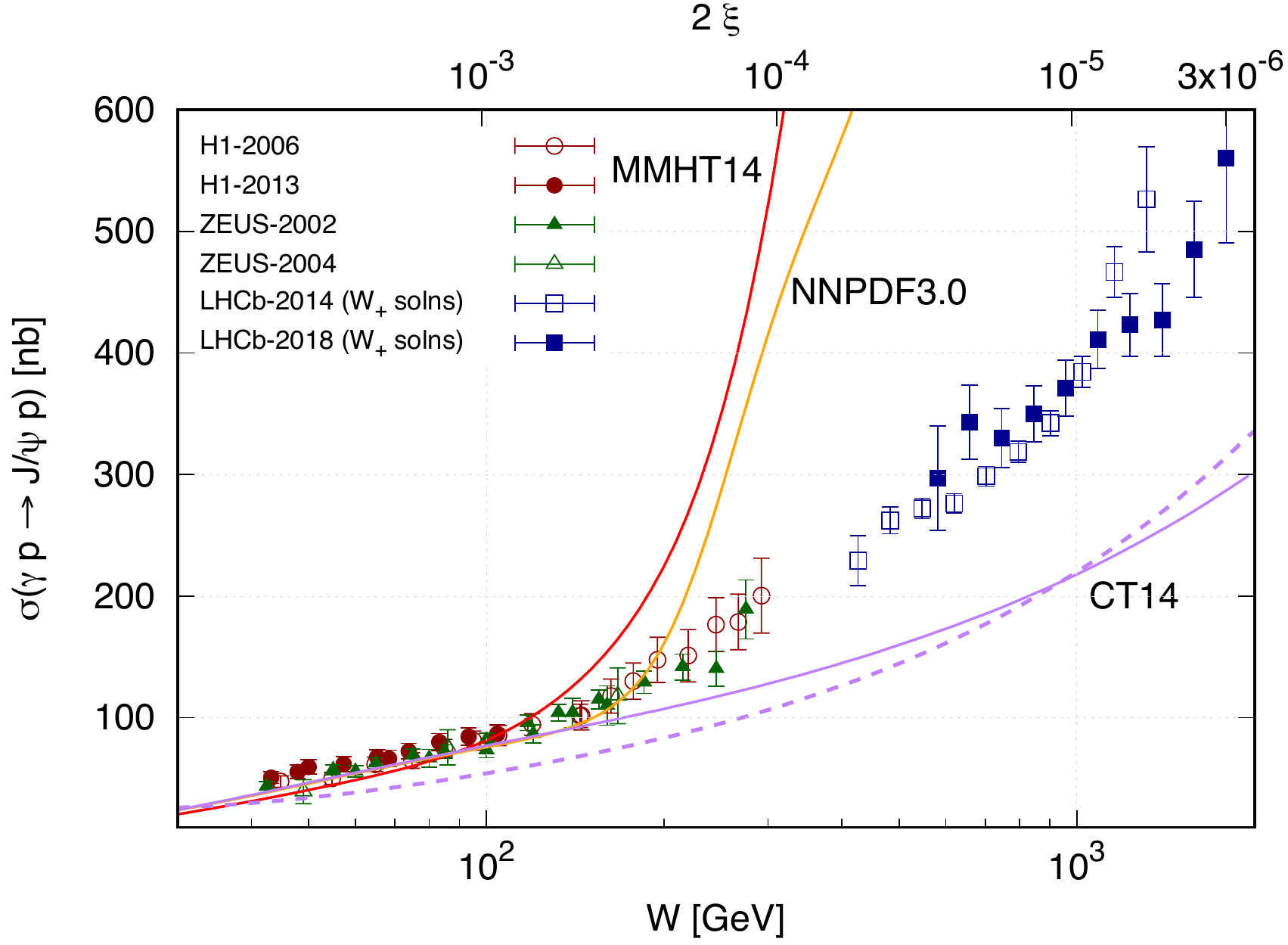}
    \qquad
    \includegraphics[scale=0.38]{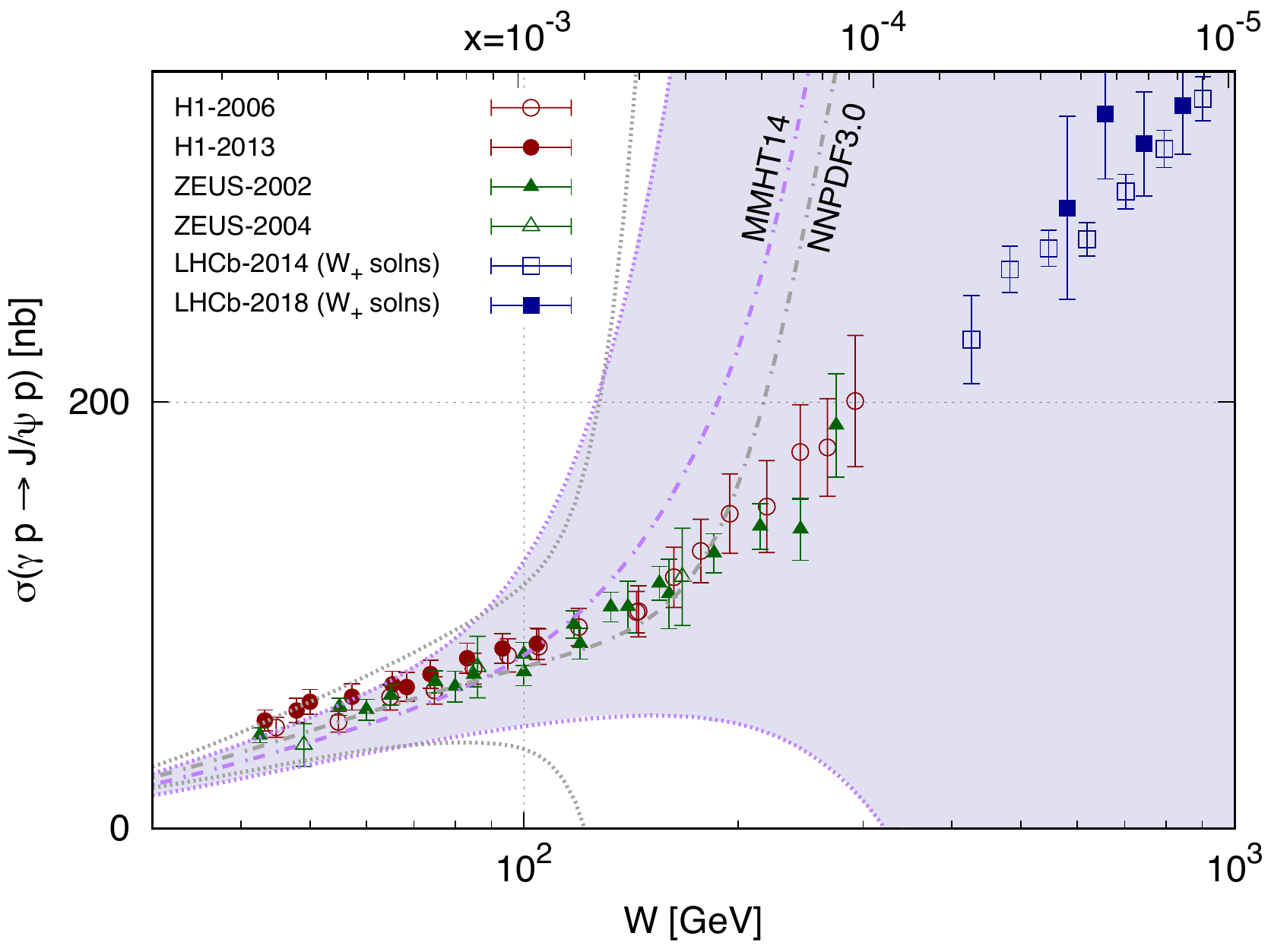} 
    \caption{ Left panel - Cross section predictions using three distinct sets of global partons~\cite{NNPDF3.0, CT14, MMHT14} with the scales $\mu_f^2 = \mu_R^2 = m_c^2$ (solid lines). Also shown for CT14 is the prediction with scales $\mu_f^2 = \mu_R^2 = 2m_c^2$ (dashed line), which demonstrates the stability of the cross section prediction with respect to scale variations. Right panel - Cross section predictions using two sets of global partons~\cite{NNPDF3.0, MMHT14} also displaying the global PDF 68\% c. l. uncertainty, which greatly exceeds the experimental uncertainty. The data are from \cite{H1ZEUS02} and the LHCb $W_+$ solutions are constructed from \cite{Ronan}, via eqn. (2.2).}
    \label{fig:my_label}
\end{figure}

\end{document}